\documentclass[default,iicol]{sn-jnl}

\jyear{2021}%

\theoremstyle{thmstyleone}%
\theoremstyle{thmstyletwo}%

\theoremstyle{thmstylethree}%

\newcommand{\norm}[1]{\left\lVert#1\right\rVert}

\begin{document}

\title[Article Title]{Machine Learning–Based Identification of Solvents from Post-Desiccation Patterns.}

\author[1]{\fnm{J. I.} \sur{Morán-Cortés}}

\author*[1]{\fnm{F.} \sur{Pacheco-V\'azquez}}\email{fpacheco@ifuap.buap.mx}

\affil[1]{Instituto de F\'isica, Benem\'erita Universidad Aut\'onoma de Puebla, Apartado Postal J-48, Puebla 72570, Mexico}


\abstract{We introduce an optimized protocol of fracture pattern classification using an artificial neural network to identify the solvent involved in the desiccation cracking process of starch-liquid slurries, even after it has been completely evaporated. For this purpose, image analysis techniques were used to characterize patterns obtained from drying suspensions using single solvents (water, ethanol, acetone) and two-component solvents (water-ethanol mixtures at different concentrations). Frequency histograms were generated based on nine morphological features, taking into account their size, shape, geometry and orientational ordering. Subsequently, we used these histograms as input data into artificial neural network variants to determine the set of features that lead to the higher accuracy in solvent identification. We obtained an average accuracy of $96(\pm 1)\%$ considering all solvents in the analysis. The highest accuracy was obtained with sets of features that include the crack area distribution. The proposed protocol can help to determine the combination of features that optimize pattern recognition in other fields of science and engineering.}

\keywords{Solvent identification, pattern recognition protocol, artificial neural networks, cracking process.}



\maketitle

\clearpage

\section{\label{sec:L1}Introduction}

In different scientific fields, neural networks stand out for their effectiveness and efficiency in automatically classifying patterns based on their morphological features \cite{2024MorehLyu,2024MashayekhbakhshMeshgini,2024WangZhao,2014PonomarevArlazarov,2016PonomarevKazanov,2014NosakaFukui,2015UmerChandraChanda,2022WangZhang,2016FanWang,2025ChoiSuh}. Ponomarev et al. \cite{2014PonomarevArlazarov,2016PonomarevKazanov} used an automatic classification method of immunofluorescent images for autoimmune disease diagnostics based on the number, size, shape, and localization of individual cells in medical tests, achieving an average accuracy of $84(\pm 6)\%$ and showing performance comparable to that of the texture-based method \cite{2014NosakaFukui}. Umer et al. \cite{2015UmerChandraChanda} proposed a person identification system based on multiscale morphology to extract structural and textural features from human iris images, with average accuracy of $99(\pm 1)\%$. Wang et al.~\cite{2022WangZhang} developed a model for material property classification based on X-ray and neutron diffraction patterns of crystal structures, reaching accuracies of $66(\pm 14)\%$ for the perovskite family $\textup{ABO}_3$. 
Fan et al. \cite{2016FanWang} propose a model for classifying multi-defect wafer map patterns by combining a densities clustering method with geometric features, including form factor (area/perimeter) and eccentricity, obtaining average accuracy of up to $94\%$, whereas more recently, Choi and Suh \cite{2025ChoiSuh} achieved over $98\%$ accuracy for the same purpose with a model focused on the active contour. Although neural networks with statistical approaches are feasible tools for pattern recognition, a correct selection of parameters is essential to avoid overfitting and increase the accuracy and efficiency of the method.

Desiccation patterns are probably the most common examples of how nature spontaneously generates configurations through complex processes that occur over multiple time and length scales. For example, at microscopic scale, slow evaporation of gold nanoparticle suspensions results in microwire patterns induced by drying liquid bridges \cite{2009VakarelskiChan}. Evaporation of blood droplets produces dendritic patterns with radial fissures in the periphery related to a high concentration of lipids \cite{2023HerreraCarreon, 2023AncheytaVelasco}. On a much larger scale, drying processes in basins and lakes generate polygonal salt patterns \cite{2023LasserNield}, while the shrinkage of frozen ground (permafrost) creates ice-filled crevices and soil \cite{2018BacchinBrutin}. Furthermore, these polygonal patterns are not limited to Earth alone, but are also found on the surface of Mars \cite{2018BacchinBrutin,2005Mangold,2019MaLowensohn,2022MontignyWalwer, 2023RapinDromart,2008PinaSaraiva} and the Moon \cite{2022MontignyWalwer}. Such patterns exhibit great morphological diversity due to the influence of multiple factors during their formation; however, they can be effectively analyzed and classified using image analysis and machine learning techniques (for example, to predict the number of cracks in the soil based on the initial moisture content, layer thickness, and sample size  \cite{2018ChoudhuryCosta}), provided that adequate selection of parameters is made.

Thin layers of colloidal suspensions have been extensively used at laboratory scale to explore desiccation processes. It has been shown that the morphology of the resulting cracking patterns depends significantly on the thickness $h$ of the layer. Theoretically, thin films crack upon reaching a critical stress, which relaxes over a distance proportional to $h$ \cite{2013Goehring}, while the crack spacing $L$ follows a power scaling with exponent $n = 2/3$ \cite{2022MontignyWalwer,1997Komatsu}, in reasonable agreement with experiments ($n=0.65\pm 0.01$ for silica particles \cite{2005AllainLimat,2004LeeRouth}, $n=0.71\pm 0.01$ for cornstarch and $n=0.53\pm 0. 02$ with clay \cite{2022MontignyWalwer}), for which the characteristic lengths \( L_1 = \sqrt{A_{\textup{c}}/N_{\textup{c}}} \) and \( L_2 = A_{\textup{c}}/L_{\textup{T}} \) have been proposed, where \( A_{\textup{c}} \) is the size of the container, \( N_{\textup{c}} \) the number of cells, and \( L_{\textup{T}} \) the total length of the cracks. In addition, the average displacement of the pattern vertices increases linearly with $h$ during repeated wetting and drying cycles \cite{2013Goehring}. Likewise, the characteristic area $A$ of the patterns is directly proportional to $h^{4/3}$ when using cornstarch with water, as well as calcium carbonate $(\textup{CaCO}_3)$ with different solvents such as water, isopropanol and silicone oil \cite{2019MaLowensohn, 2022MontignyWalwer}. Differences in the fitting parameters are evidence of the influence of the solute and solvent on the size of the patterns.

In order to classify fracture patterns, frequency histograms are a useful tool to represent their morphological distribution. A variety of parameters exists to characterize the geometry and topology of these patterns based on radial density profile \cite{2023HerreraCarreon}, average areas \cite{2019MaLowensohn,2009PauchardAbou,2011ChaoYu, 2017AkibaMagome,2022RoyHaque}, lengths \cite{2022MontignyWalwer,2008PinaSaraiva,2005AllainLimat,2004LeeRouth,2011ChaoYu,2005BohnPlatkiewicz}, number of neighbors \cite{2008PinaSaraiva,2022RoyHaque,2005BohnPauchard}, fractal dimension \cite{2014DeCarloShokri,2009BaerKent} and lacunarity \cite{2009BaerKent}, to name a few, but only some studies include a large dataset of measurements. However, histograms based on the normalized isoperimetric ratio, defined as \(\lambda = 4\pi A/P^2\), where \(A\) is the area and \(P\) is the perimeter of the patterns, have been proposed by Roy et al. \cite{2022RoyHaque} as the most efficient parameter to distinguish between fracture patterns, which they qualitatively evidenced in cracked materials of natural mud, clays (such as laponite and bentonite), corn and potato starch extracts, and polymers (such as PDMS), among others. From a statistical point of view, this should hold true even if the successive cleavage of the fracture patterns does not follow a regular or uniform process \cite{2005BohnPauchard}. For this reason, histograms of the area$-$perimeter set and other morphological features may contribute to the automated identification of such patterns, but the effectiveness and efficiency of these features for classification is still unknown.

In this work, we propose an optimized protocol to use neural networks for the recognition and classification of fracture patterns produced during the desiccation of cornstarch-liquid slurries involving different solvents. For this purpose, an experimental array (described in Sec. \ref{sec:L2}) was used to follow the evaporation process of thin layers of the slurries until the fracture patterns appear. In Sec. \ref{sec:L3}, we describe the methodology used for the image analysis. Then, we propose a set of features represented in frequency histograms to characterize fracture patterns from their size, shape, geometry and orientational ordering. On the one hand, size is determined through the area and perimeter of the cells and the area of the cracks, all normalized by characteristic values, while shape is described by the eccentricity. On the other hand, the geometry is determined by the number of neighbors, obtained from Voronoi diagrams; while ordering is determined by an order parameter, defined as a function of these neighbors and the centroids of the cells. In Sec. \ref{sec:L4}, we outline the framework of the proposed protocol, emphasizing the construction, training, and accuracy calculation of an Artificial Neural Network (ANN) to classify experimental sample classes. Subsequently, in Sec. \ref{sec:L5}, we evaluated the accuracy of the classification, considering each individual feature, as well as combinations of different sets of features. Finally, the conclusions are presented in Sec. \ref{sec:L6} to summarize the most relevant aspects of this work.

\section{\label{sec:L2}Experimental setup}

\begin{figure}[h!]
    \centering
    \includegraphics[width=1.0\linewidth]{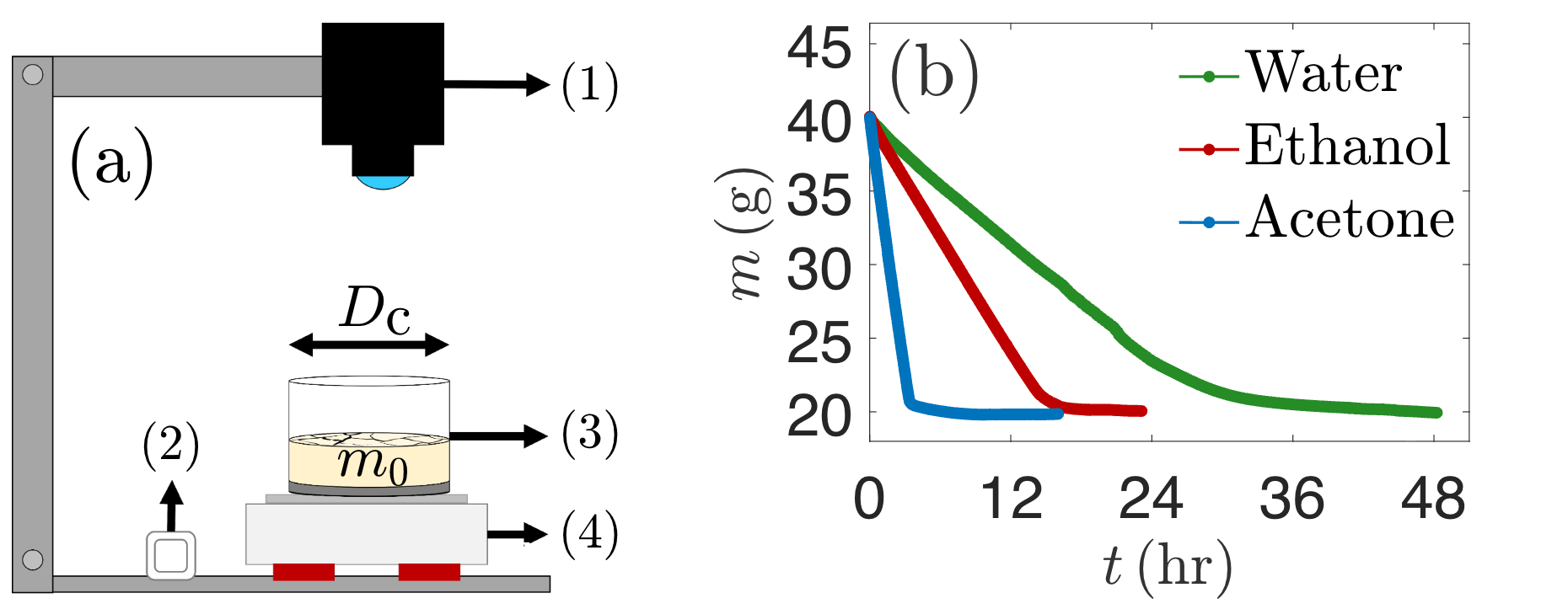}
    \caption{(a) Experimental setup components: (1) digital camera, (2) digital thermometer-hygrometer (3) aluminum container with initial mass $m_0$ of a specific mixture, and (4) analytical balance. (b) Remanent mass $m$ of the mixtures as a function of time $t$ during the drying process, for three different solvents.}
    \label{Fig1}
\end{figure}

The experimental design is sketched in Fig. \ref{Fig1}a, and it is similar to previous systems used in the literature \cite{2009PauchardAbou,2011ChaoYu}. A given mass of cornstarch-liquid slurry is deposited in a cylindrical aluminium container of diameter $D_{\textup{c}} = 100$ mm,  and the container is placed on an analytical balance (OHAUS $\textup{PX}423$).  The system allows for recording the mass of the slurry during the desiccation process every minute, and a digital thermometer-hygrometer placed inside the balance chamber was used to register the temperature $T\left(^\circ\textup{C}\right)$ and the relative ambient humidity $\textup{RH}\left(\%\right)$ every five minutes. A digital camera (NIKON $\textup{D}3200$) is positioned to record images from a top view of the container, and to capture the final morphology of the fracture patterns.

The mixtures were prepared with an initial concentration $\phi_0 = 1/2$, defined as $\phi_0 =m_{\small\textup{liq}}/\left(m_{\small\textup{liq}}+m_{\small\textup{cs}}\right)$ \cite{2012CrostackNellesen}, where $m_{\small\textup{liq}}$ is the mass of the liquid (solvent) and $m_{\small\textup{cs}}$ corresponds to dry cornstarch (solute), obtained from a commercial brand (Maizena). A mass $m_0 = 40.0\left(\pm 0.1\right)$g of the homogeneous mixture was prepared for each experiment using equal masses of liquid ($20.0$g) and cornstarch ($20.0$g), obtaining average densities $\rho = 1.20(\pm 0.01)\,\textup{g}/\textup{cm}^3$ for cornstarch-water slurry and $\rho = 1.03 (\pm 0.01) \,\textup{g}/\textup{cm}^3$ for both cornstarch-ethanol and cornstarch-acetone slurries. The mixture is deposited in the center of the container, and it extends radially, forming a thin horizontal liquid layer of thickness $h_0 \sim 4-5 ~\rm{mm}~\ll D_{\textup{c}}$. Once the solvent evaporation is completed, the layer thickness (and therefore, the volume of the layer) is reduced by $16(\pm 1)\%$, and the patterns are recorded for subsequent analysis.

Following the procedure described above, 75 liquid layers were prepared using single solvents, distributed in 25 samples for each solvent: water ($\gamma = 72.01\,\textup{mN}/\textup{m}$), ethanol ($\gamma = 21.82\,\textup{mN}/\textup{m}$), and acetone ($\gamma = 23.02\,\textup{mN}/\textup{m}$), where $\gamma$ is the surface tension at $25^\circ\textup{C}$~\cite{1995VazquezAlvarez,2007EndersKahl}. The samples were left to evaporate naturally for 48, 24, and 12 hours for the water, ethanol, and acetone mixtures, respectively. The latter substances are more volatile and require less time for total evaporation, which is considered to occur when $m\approx 20$ g, i.e. when all the liquid has been evaporated and only the solute remains in the container.
Fig. \ref{Fig1}b shows how $m$ decreases with time $t$ for the three samples. Environmental conditions remained nearly constant throughout the experiments, with an average relative humidity of $\textup{RH}_{\textup{env}} = 39 (\pm 6)\%$ and an average room temperature of $T_{\textup{env}} = 28 (\pm 1)^\circ\textup{C}$. From the moment of deposition, evaporation proceeds in two stages \cite{2009PauchardAbou,2006GoehringMorris,2009Goehring}. First, the mass decreases linearly with time (constant evaporation rate) until a wet cornstarch layer seems to settle at the bottom of the container, which is characterized by the notorious change of curvature in the plots. At this moment, the layer adheres to the substrate and begins to contract until it detaches from the container walls. This process is consistent for different solutes in water, such as cornstarch \cite{2019MaLowensohn,2005BohnPlatkiewicz}, granular polymers \cite{2024LiaudatMuraro}, Romainville clay \cite{2011ChaoYu}, hard and soft nanolatex spheres \cite{2009PauchardAbou,2005BohnPauchard}, Sisal fiber and soil \cite{2024BuLiu}, sand and kaolinite clay \cite{2014DeCarloShokri}, among others. Subsequently, two types of cracks appear that divide the layer into primary and secondary patterns according to their order of appearance \cite{2019MaLowensohn,2005BohnPlatkiewicz}. Primary patterns arise from a critical $h_{\textup{c}}$ layer thickness and they are formed in the first evaporation stage due to capillary contraction of the material \cite{2019MaLowensohn}. During the second evaporation stage for the case of cornstarch in water, secondary patterns emerge in large amounts and simultaneously within the primary patterns due to dehydration of the hygroscopic cornstarch particles \cite{2019MaLowensohn,2005BohnPauchard}, even when the layer thickness is $h < h_{\textup{c}}$.

\begin{figure}[h!]
    \centering
    \includegraphics[width=1.0\linewidth]{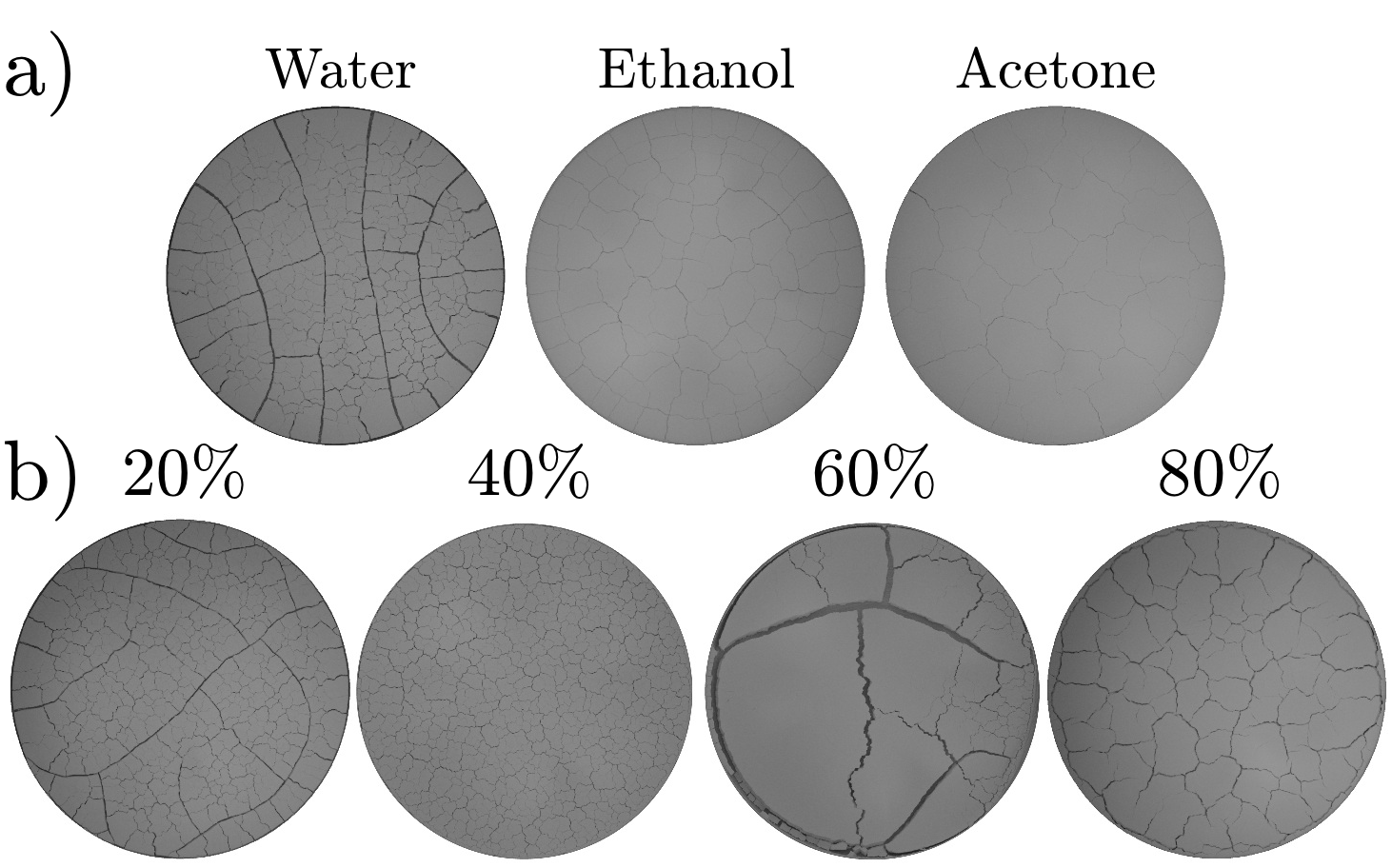}
    \caption{Fracture patterns observed in experimental samples of mixtures using solvents of a) pure liquids and b) water-ethanol mixtures with the indicated percentages of ethanol. }
    \label{Fig2}
\end{figure}

Figure \ref{Fig2} shows that the final fracture patterns depend on the solvent used. For the case of water, primary thicker fractures are clearly observed (secant lines  crossing almost the whole layer), which are not observed with ethanol or acetone. There are also numerous small secondary thinner fractures associated with columnar joints in water  \cite{2013Goehring,2006GoehringMorris,2009Goehring}, while the number of cells decreases significantly in ethanol and acetone, causing an increase in their size, as shown in Fig. \ref{Fig2}a. Furthermore, the secondary cells are larger in acetone than in ethanol, whereas in water they are noticeably smaller and more irregular. Consequently, the samples in Fig. \ref{Fig2}a can in principle be easily distinguished from one another.

\begin{table}[h!]
\tiny
\centering
\setlength{\tabcolsep}{4pt} 
\renewcommand{\arraystretch}{1.0}
\begin{tabular}{|c|c|c|c|c|c|c|}
\hline
\begin{tabular}[c]{@{}l@{}} Ethanol content \\ in water\end{tabular}& 0\% & 20\% & 40\% & 60\% & 80\% & 100\% \\ \hline \hline
$X$ &
$0.000$ & $0.089$ & $0.207$ & $0.370$ & $0.610$ & $1.000$ \\ \hline
$\gamma$ (mN/m) &
$72.01$ & $37.97$ & $30.16$ & $26.23$ & $23.82$ & $21.82$\\ \hline 
\end{tabular}
\vspace{0.2cm}
\caption{Ethanol-water binary solvents. Mole fraction concentration $X$ of ethanol $(M = 46.07)$ in water ($M = 18.02$) depending on the indicated percentage of ethanol volume in the mixture, where $M$ is the molar mass.  The  corresponding surface tension $\gamma$ at $25^\circ$C is indicated \cite{1995VazquezAlvarez,2001PolingPrausnitz,2020CervantesEscobar}.}
\label{Tab1}
\end{table}

The change in pattern features can be associated to changes of surface tension \cite{2013Goehring}. To explore this aspect, additional 50 liquid layers with binary water-ethanol solvents were considered at different concentrations to decrease systematically the surface tension of the solvent (see Table \ref{Tab1}). The characteristic patterns of the transition from pure water to pure ethanol are shown in Fig. \ref{Fig2}b using water-ethanol mixtures at 20, 40, 60 and 80 $\%$ of ethanol content. Some morphological features of the dominant liquid (present in greater proportion) are retained, while those of the recessive liquid (present in smaller proportion) are difficult to perceive visually with the fracture pattern. On the one hand, for concentrations close to pure liquids, the number of patterns and the size of the cracks appear similar to those of the dominant liquid, although contrasting morphological features arise due to the recessive liquid. On the other hand, mixtures with comparable proportions of water and ethanol can be clearly distinguished from the other samples. For instance, at $40\%$ ethanol, primary cracks disappear with respect to $20\%$, but the number of secondary cells looks similar. A particular pattern appears at $60\%$ ethanol, in which case the pattern size and width of the cracks seem considerably larger, probably due to the exothermic reaction of the mixture that generates an increase in temperature and dilation of the slurry during the desiccation process.  However, in the mixture of $80\%$ ethanol, a clear competition arises between the morphological features obtained with pure liquids; the presence of water produces thicker fractures, but with an ordering and geometry of cells similar to those observed in pure ethanol. Therefore, the number of patterns, the spatial distribution, their average area and perimeter, and the thickness of the cracks are morphological features of the patterns determined by the solvent and should be considered in the classification.

\section{\label{sec:L3}Morphological analysis}

\begin{figure}[h!]
    \centering
    \includegraphics[width=1.0\linewidth]{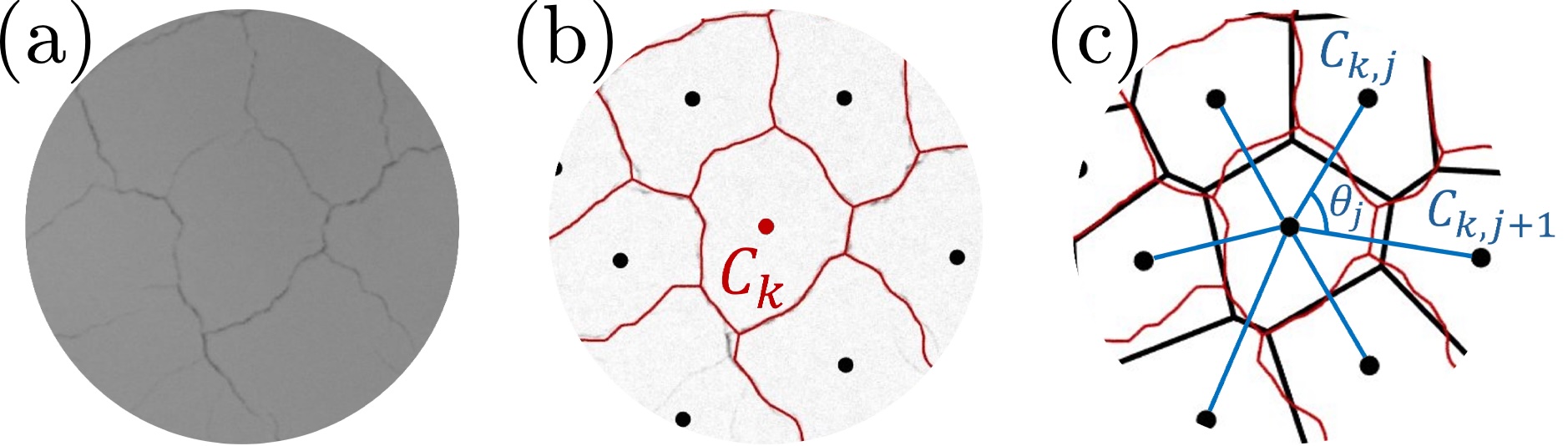}
    \caption{Image processing. (a) Picture of a fracture pattern. (b) Corresponding centroids (black dots) and (c) Voronoi diagram (black edges) from the real fracture pattern (red edges) after image processing.}
    \label{Fig3}
\end{figure}

Figure \ref{Fig3} illustrates the methodology used for the morphological analysis of fracture patterns. Images of all experimental samples are first captured, as illustrated in Fig. \ref{Fig3}a. Then, using the software FIJI ImageJ\textsuperscript{\textregistered}, each image is binarized to show exclusively the cracks that outline the boundaries of the patterns as depicted by the red borders in Fig. \ref{Fig3}b. By performing this image analysis iteratively for each of the experimental samples, five main features can be estimated: the number $N_{\textup{c}}$ of patterns in each experimental sample, the area $A$, perimeter $P$, centroid $C\left(x,y\right)$, and eccentricity $\varepsilon$ of the patterns. After applying Voronoi tesselation, the total length $L_{\textup{T}}$ was obtained. Furthermore, each image was divided in $n=64$ circular sectors to account for variations in the crack distribution, and the number of pixels conforming the cracks in each sector was estimated, which allows to obtain the area $A_{\textup{cr}}$ of the cracks. These morphological features characterize the size (area$-$perimeter) and shape (eccentricity) of the fracture patterns.

\begin{figure*}[h!]
    \centering
    \includegraphics[width=155mm]{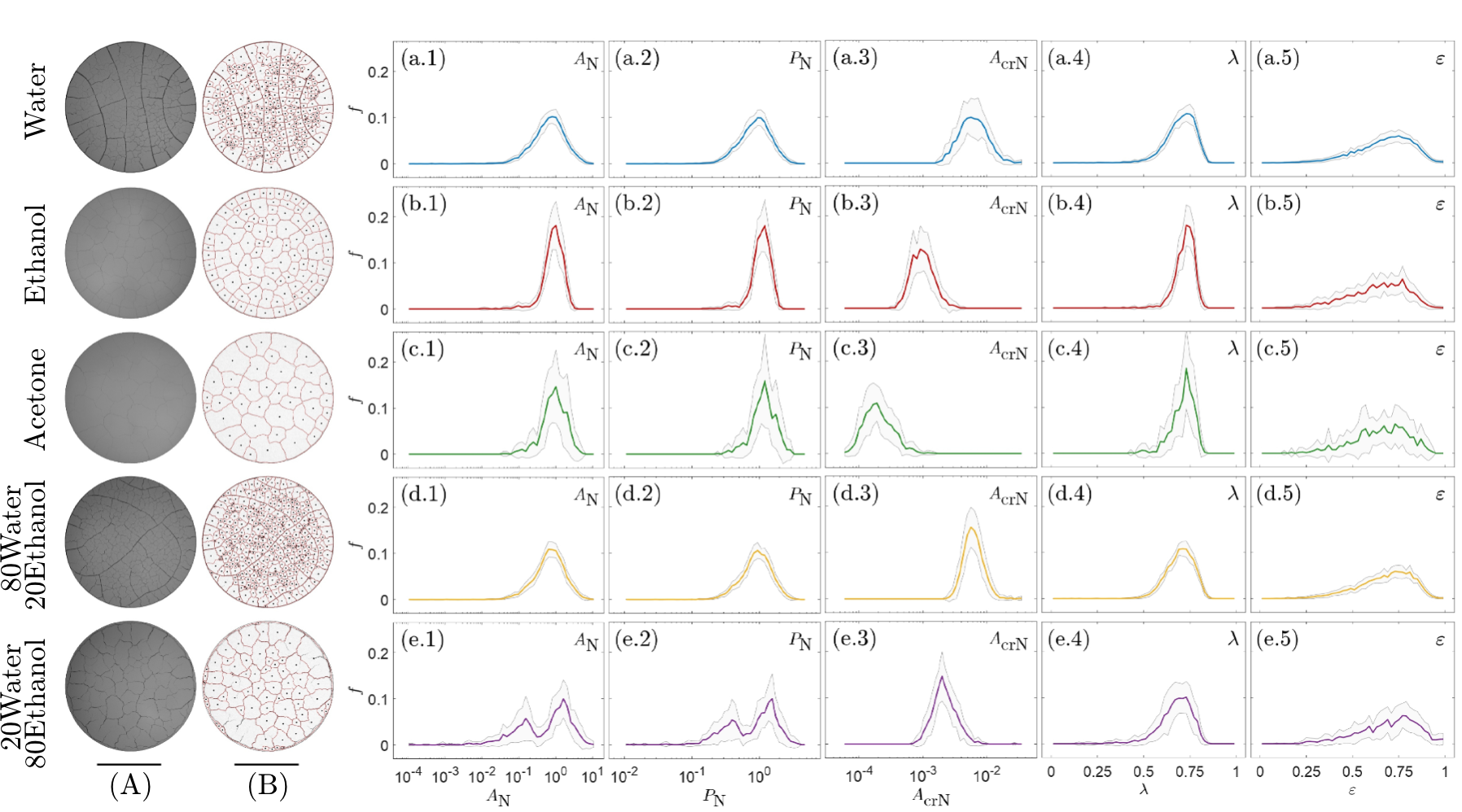}
    \caption{Size-shape analysis. (A) Final patterns observed after the desiccation process of layers of cornstarch and the indicated solvent. (B) Corresponding processed images used to identify the cracks (red edges) and their centroids (black dots). In A and B the scale bar is $50\,\textup{mm}$. (a-e) Histogram distribution of (1) pattern area $A_{\textup{N}}$, (2) perimeter $P_{\textup{N}}$, (3) crack area $A_{\textup{crN}}$, (4) isoperimetric ratio $\lambda$, and (5) eccentricity $\varepsilon$, obtained using as solvent: (a) water, (b) ethanol, (c) acetone, and (d) $80/20\%$ and (e) $20/80\%$ water/ethanol. Colored lines represent the average value and the gray shaded area represents the standard deviation. }
\label{Fig4}
\end{figure*} 

On the other hand, centroids (Fig. \ref{Fig3}b) allow to infer geometrical and ordering features of the patterns, which show a high correspondence with the Voronoi diagrams (black edges in Fig. \ref{Fig3}c) \cite{2022RoyHaque}. Using MATLAB\textsuperscript{\textregistered}, if we focus on the Voronoi diagram of the centroid \( C_k \) of some pattern (see Fig. \ref{Fig3}b), we can compute its number of neighbors $N_{\textup{B},k}$ with the associated Voronoi diagram to know its geometry. In addition, such centroid can be linked to two adjacent and consecutive centroids \( C_{k,j} \) and \( C_{k,j+1} \) by straight lines (blue lines in Fig. \ref{Fig3}c) to calculate the angle \( \theta_j \) between them. Therefore, the associated bound orientational parameter (BOP) \( \psi_{k,n} \) of each $k-$th pattern \cite{2023LopezPacheco}, defined as a function of the number $N_{\textup{B},k}$ and the angles \( \theta_j \) for each \( j = 1, \dots, N_{\textup{B},k} \), can be computed by: 
\begin{equation}
    \psi_{k,n} = \frac{1}{N_{\textup{B},k}} \norm{  \sum_{j=1}^{N_{\textup{B},k}} \exp{\left(in\theta_j\right)} }, \hspace{0.4cm} 0\leq \theta_j<2\pi 
\end{equation}
with
\begin{equation}
    \theta_j = \arccos\left[\frac{\left(C_{k,j+1}-C_k\right)\cdot \left(C_{k,j}-C_k\right)}{\norm{C_{k,j+1}-C_k}\norm{C_{k,j}-C_k}}\right].
\end{equation} 

By applying the morphological analysis to all fracture patterns, we initially estimate the average values of such morphological features. On the one hand, since the area of the container $A_{\textup{c}}=7854$ mm$^2$ is constant, then the average area $\overline A$ of the patterns is inversely proportional to $\overline N_{\textup{c}}$, such that: $\overline A \approx A_{\textup{c}}/\overline N_{\textup{c}} =L_1^2$. On the other hand, the measurements obtained from the image analysis also reveal that $\overline P\approx 8 A_{\textup{c}}/ \overline L_T = 8L_2$. Additionally, the patterns exhibit similar characteristics; they follow the normalized isoperimetric ratio with an average value of $\lambda = 0.70\pm 0.03$, and they show comparable average values of eccentricity $(\varepsilon = 0.67\pm 0.07)$ and order parameter $(\psi = 0.47\pm 0.17)$. Therefore, normalized quantities for the sizes of the patterns and cracks can be defined relative to these characteristic lengths, i.e., $A_{\textup{N}} = A/L_1^2$, $P_{\textup{N}} = P/\left(8 L_2\right)$, and $A_{\textup{cr}N} = A_{\textup{cr}}/L_3^2$, respectively, where $L_3 \equiv L_1^2 / L_2=L_T/N_{\textup{c}}$. Then, a set of normalized features can be used for the morphological characterization of the fracture patterns. 


Figure \ref{Fig4} displays the size and shape measurements in the form of frequency histograms of normalized features, which help to identify those parameters that can be useful to classify the experimental samples. To obtain the histograms, the methodology described above was iteratively applied to the samples (see Figs. \ref{Fig4}A-B), with the goal of extracting and characterizing nearly $24,000$ individual cells. In the histograms, one can notice that the measurements for ethanol are more homogeneous than for the case of water (which shows a wider distribution). For the mixture of $20\%$ water and $80\%$ ethanol, two maxima appear in the distributions of $A_{\textup{N}}$ and $P_{\textup{N}}$ indicating a superposition of the effects of both solvents. There is also a noticeable shift to the left in the maximum of the $A_{\textup{cr}N}$ distribution for more volatile liquids. This indicates thinner cracks for acetone and ethanol compared to water. For the binary mixtures, the maximum is between the values corresponding to the pure solvents, and the shift depends on the ethanol concentration, suggesting that $A_{\textup{cr}N}$ could be one of the most relevant parameters for the classification.

\begin{figure*}[h!]
    \centering
    \includegraphics[width=155mm]{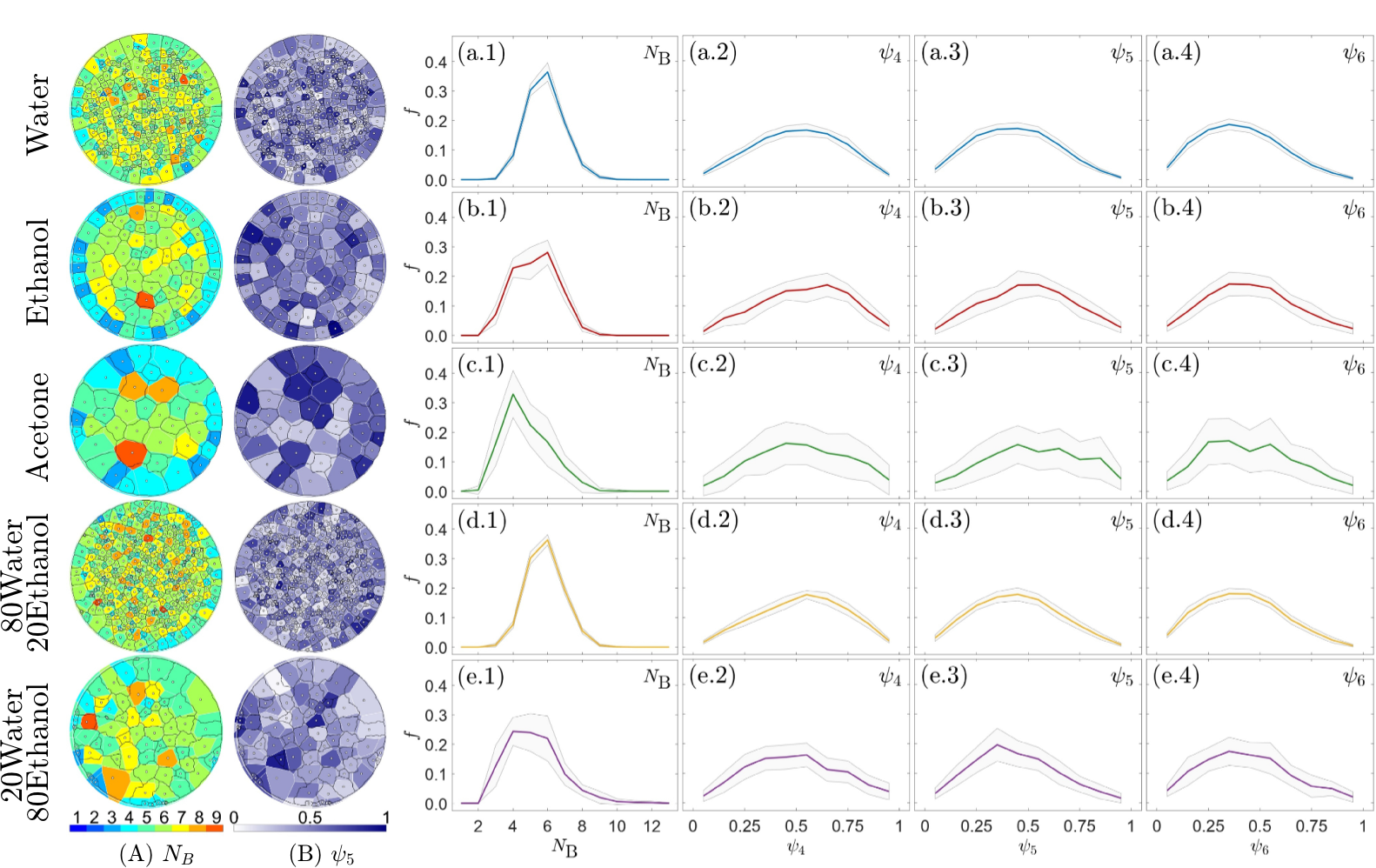}
    \caption{Orientational ordering analysis. (A-B) Voronoi diagrams (colored cells) and centroids (white dots) obtained from the actual patterns (black lines) depending on the solvent. Colors are used to indicate A) the number of neighbors $N_B$ and B) the orientational order (here $\psi_5$ is shown for visualization) according to the color bars. (1-4) Frequency histograms of 1) nearest neighbors, and different order parameters 2) $\psi_4$, 3) $\psi_5$ and 4) $\psi_6$, depending on the solvent indicated in the same row. Colored lines represent the mean values and the gray shaded areas the corresponding standard deviations.}
    \label{Fig5}
\end{figure*}

Figure \ref{Fig5} shows histograms corresponding to ordering parameters ($\psi_4, \psi_5, \psi_6$) and number of neighbors $N_B$, where it is more difficult to distinguish by eye the differences between the distributions. The histograms show that the patterns have between 4 and 7 neighbors, while on average they tend toward hexagonal geometries $(N_{\textup{B}}=6\pm 1)$ in samples with pure water and, $80\%$ water and $20\%$ ethanol mixture, but they tend toward pentagonal geometries $(N_{\textup{B}}=5\pm 1)$ in those with pure ethanol, pure acetone and, $20\%$ water and $80\%$ ethanol mixture (see Fig. \ref{Fig5}A). On the other hand, the order parameter is used to analyze whether the centroids of the patterns are organized in an ordered $(\textup{BOP} \to 1)$ or disordered $(\textup{BOP} \to 0)$, but from Figs. \ref{Fig5}(2$-$4) one finds that the $47\left(\pm 2\right)\%$ of the patterns present values of $\textup{BOP}$ between $0.35$ and $0.65$ (see Fig. \ref{Fig5}B), which are around their mean value; therefore, it is concluded that the centroids do not manifest a predominant symmetry associated with square, pentagonal or hexagonal structures. 

\section{\label{sec:L4}Deep learning implementation}

\begin{figure*}[h!]
    \centering
    \includegraphics[width=155mm,trim=25mm 25mm 25mm 25mm,clip]{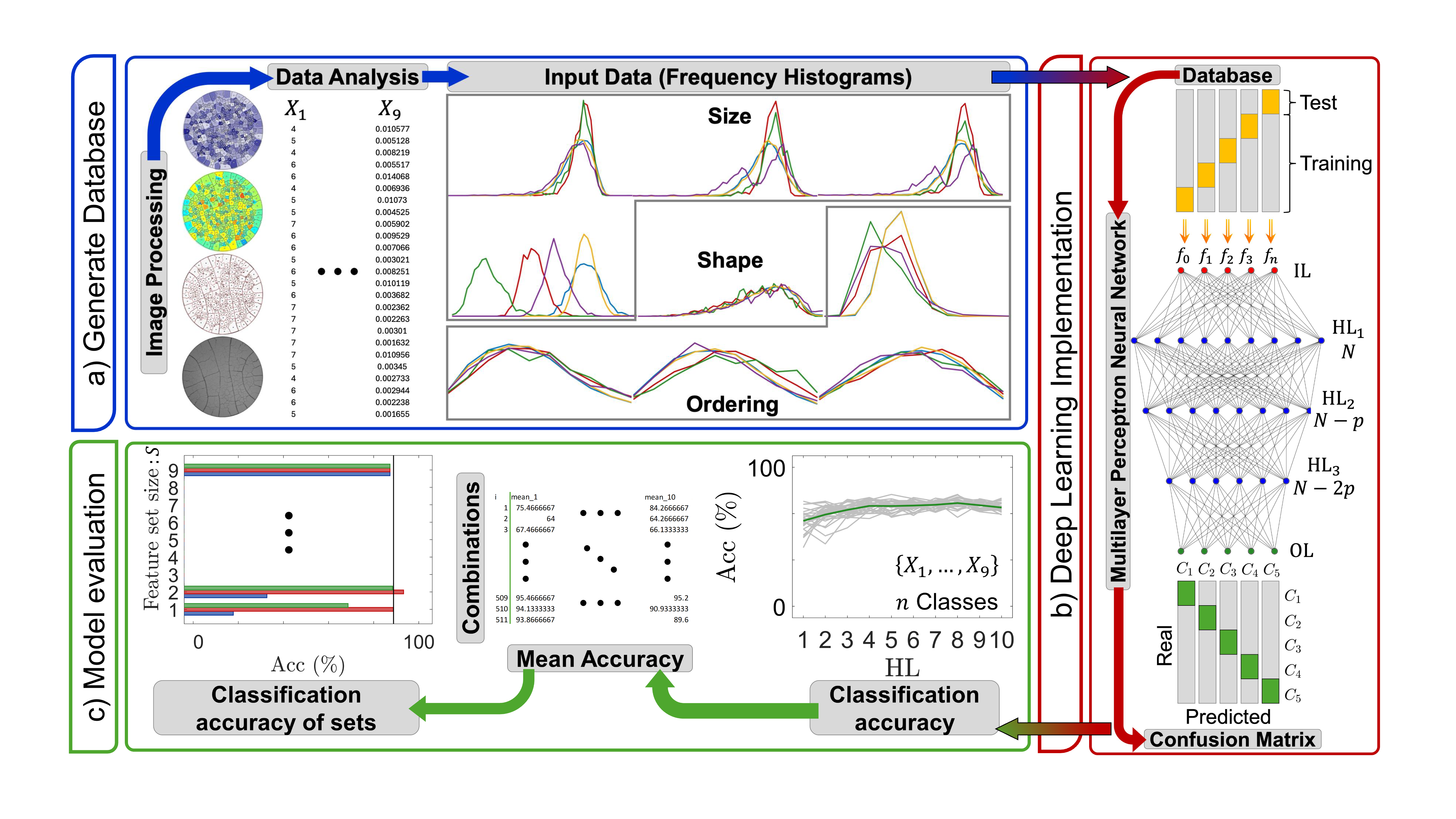}
    \caption{Framework of the proposed protocol, including: (a) database construction, (b) machine learning implementation, and (c) model evaluation.}
    \label{Fig6}
\end{figure*}

Let us now describe the deep learning approach used to identify the solvent involved in the desiccation process (see framework in Fig. \ref{Fig6}). First, we apply the fracture pattern analysis described in the previous section to obtain the frequency histograms (Fig. \ref{Fig6}a). Then, the Artificial Neural Networks (ANNs) is trained in MATLAB\textsuperscript{\textregistered} using a 5-fold cross-validation (Fig. \ref{Fig6}b), where the dataset is divided into five equal parts: four of them $(80\%)$ are used as the training set, and the remaining one $(20\%)$ as a test set (shown in yellow), and the latter is rotated until all five possible combinations are covered \cite{2025ChoiSuh,2023KimBharath}. Then, a multilayer perceptron (MLP) neural network with a given number of neurons $N$ is employed, which consists of fully connected layers: an input layer ($\textup{IL}$), one or more hidden layers ($\textup{HL}=1-10$), and an output layer ($\textup{OL}$). Here, the frequency histograms are used as input data $(f_1,f_2,\ldots,f_n)$ and the different solvents are defined as output data $(C_1,\ldots,C_5)$. Subsequently, the neural network is trained using the Scaled Conjugate Gradient (SCG) backpropagation algorithm, which iteratively adjusts weights and biases to minimize errors, with three predefined random seeds to ensure the reproducibility. For each hidden layer, the hyperbolic tangent sigmoid function $\textup{tanh}(x) =  2/(1+\exp(-2x))-1$ is used as the activation function, while the softmax function $y_i =\exp(x_i)/\left[\sum_{j=1}^N\exp(x_j)\right] $ is assigned to the output layer.  Using an initial number of neurons $N$, ranging from $60$ to $120$ in increments of $2$ neurons, 31 neural network variants are evaluated for each individual feature $\{X_1,\ldots,X_9 \}$ across ten hidden layers with steps of $p=6$ neurons. The confusion matrix highlights the correctly classified samples (shown in green), which vary for each neural network variant, and the classification accuracy is calculated as the ratio of the number of correct predictions to the total number of tests. From 465 runs per hidden layer (5 folds $\times$ 3 seeds $\times$ 31 variants), an average accuracy curve is obtained, illustrated by the green curve of the $\textup{Acc}$ vs $\textup{HL}$ plot in Fig. \ref{Fig6}c. This process is performed with all the size, shape, and ordering features to determine which of them lead to a higher classification accuracy, individually or forming feature sets of different size $(S=1,\ldots,9)$.

\section{\label{sec:L5}Model Evaluation}

\begin{figure*}[h!]
    \centering
    \includegraphics[width=0.9\linewidth]{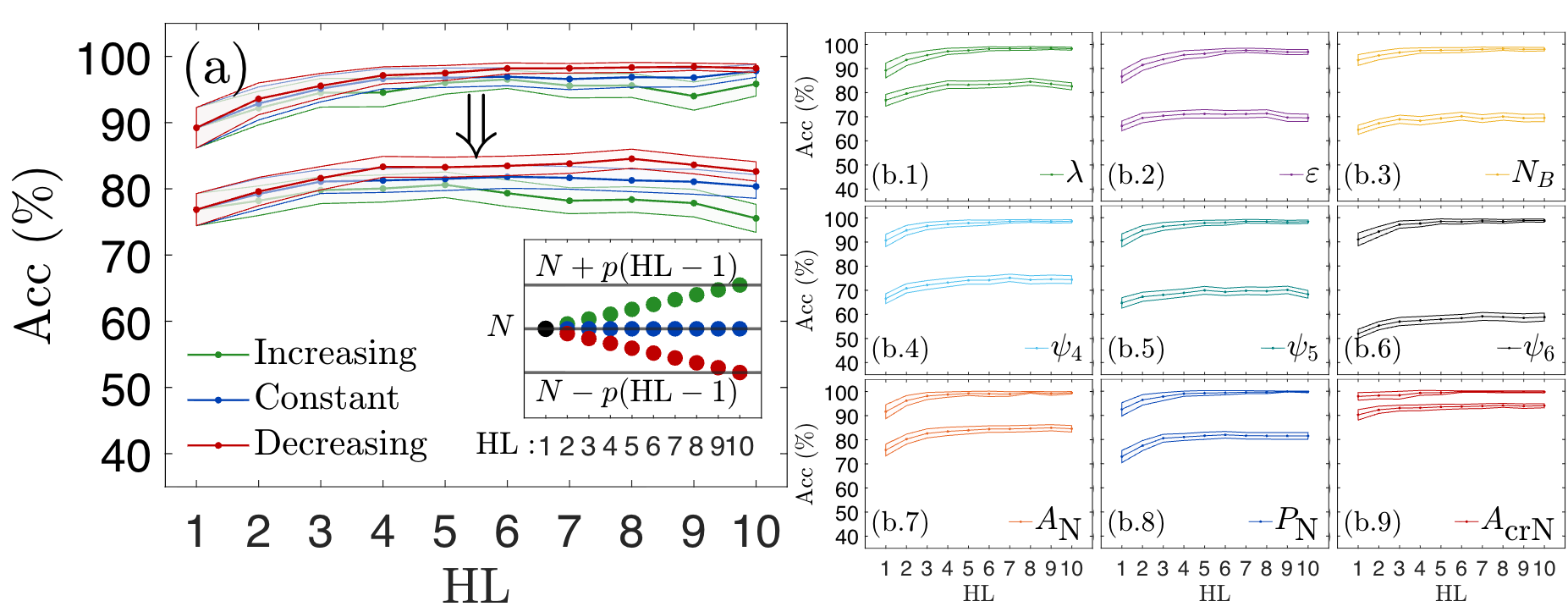}
    \caption{Classification accuracy. Performance of (a) the normalized isoperimetric ratio $\lambda$ in three MLP neural network configurations depicted in the inset: increasing, constant, and decreasing architectures as a function of the number of hidden layers (HL). (b) Performance of each feature in the decreasing architecture. Solid lines indicate mean values, and shaded areas represent the corresponding standard deviations.}
    \label{Fig7}
\end{figure*}

Figure \ref{Fig7} shows the variation in the accuracy of the classification as a function of the number of hidden layers. The upper curves (higher accuracy) in each plot correspond only to the classification of single solvents (water, ethanol, and acetone), while the bottom curves (lower accuracy) also include binary solvents. In particular, using the isoperimetric ratio, Fig. \ref{Fig7}a exemplifies the accuracy reached  for three MLP neural network configurations obtained by decreasing, keeping constant, or increasing the number of neurons in each hidden layer (as shown in the inset). One can notice that the decreasing MLP neural network achieves the highest accuracy (red symbols in Fig. \ref{Fig7}a). Thus, we can improve the accuracy of the classification by choosing this architecture, which is the one used in the rest of this study.

Figures \ref{Fig7}c present a comparison of the classification accuracy achieved using nine individual features $\left(\lambda,\varepsilon,N_{\textup{B}},\psi_4,\psi_5,\psi_6,A_N,P_N,A_{\textup{cr}N}\right)$ as input to the 31 variants of the decreasing MLP neural network, in order to identify those parameters contributing to higher or lower predictive performance. When all the solvents are included (lower curves), the order parameter $\psi_6$ yields the lowest performance, with accuracies ranging from $51(\pm 2)\%$ to $59(\pm 2)\%$; $\varepsilon$,$N_{\textup{B}}$, $\psi_4$ and $\psi_5$ achieve values from $64(\pm 2)\%$ to $75(\pm 2)\%$. In comparison, the normalized parameters $\lambda$, $P_N$ and $A_N$ reach higher values from $73(\pm 3)\%$ to $85(\pm 1)\%$. Notably, the normalized crack area $A_{\textup{cr}N}$ yields the highest performance, ranging from $90(\pm 2)\%$ to $94(\pm 1)\%$; therefore, this is a key parameter to maximize the accuracy of the classification.

\begin{figure*}[h!]
    \centering
    \includegraphics[width=0.9\linewidth]{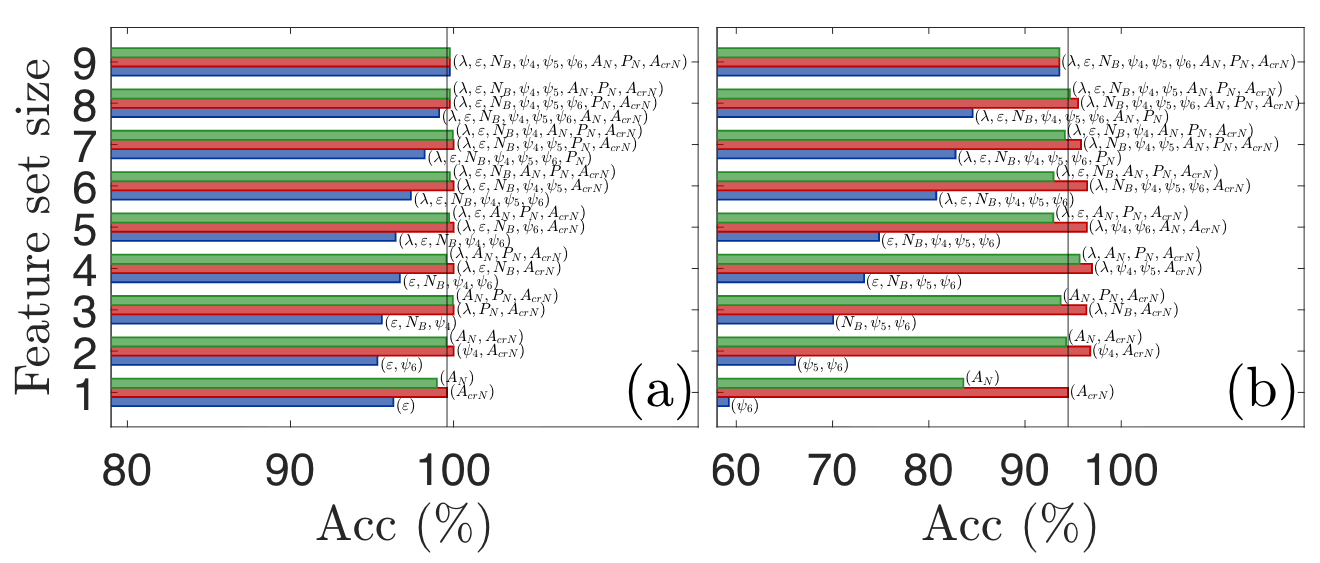}
    \caption{Classification accuracy for some sets of features (with $N=60$ neurons), comparing (a) three classes corresponding to single solvents (water, ethanol, and acetone) and (b) five classes including the single solvents and binary solvents ($20\%$ and $80\%$ ethanol content in water). For sets with the same number of elements, the worst, intermediate and best accuracies are indicated by blue, green and red bars, respectively. For comparison, green bars correspond to the same sets in both cases (a) and (b). The vertical black line corresponds to the accuracy using only $A_{\textup{cr}N}$, which leads to the highest value obtained with an individual feature.}
    \label{Fig8}
\end{figure*}

The combination of features can be used to improve even more the accuracy, using concatenated histograms as input data of the decreasing MLP neural network. By combining the features, there are $2^9-1=511$ possible sets with at least one feature. Thus, we calculated the average accuracy of each combination to identify the set of features that maximizes the accuracy, and avoid sets with overfitting that decrease its overall value. 
Figure \ref{Fig8} presents those sets with the lowest (blue), intermediate (green) and highest (red) average accuracies, depending on the number of parameters conforming the set. For the case of single solvents (Fig. \ref{Fig8}a), the combination of parameters leads to very high accuracies $>95\%$. Even with a two-feature set \(\left(\psi_4,A_{\textup{crN}}\right)\), the highest accuracy is reached ($100.0\%$), but using the full nine-features set results in overfitting, with a slight decrease of accuracy and inconvenient longer processing time. On the other hand, including binary solvents in the analysis (see Fig. \ref{Fig8}b), the classification is certainly less accurate. In this case, the two-feature set \(\left(\psi_5,\psi_6\right)\) yields the lowest classification performance, with an accuracy of \(66.1(\pm 2.8)\%\). In contrast, the four-feature set \(\left(\lambda,\psi_4,\psi_5,A_{\textup{crN}}\right)\) yields the highest accuracy of \(97.0(\pm 1.2)\%\), slightly better than the two-feature set \(\left(\psi_4,A_{\textup{crN}}\right)\), which accuracy is \(96.8(\pm 1.4)\%\). The use of all the parameters also results in overfitting in this case, with an accuracy of \(93.6(\pm 1.6)\%\). Therefore, the most suitable set of features for the classification in both groups of solvents is \(\left(\psi_4,A_{\textup{crN}}\right)\), associated with the orientational ordering and the area of the cracks. It is important to remark that the measurement and subsequent analysis of data is simpler for those features 
that can be easier to determine. For instance, an adequate feature set to analyze unknown samples of fractured dried layers that requires less processing time is the set associated with size $\left(\lambda, A_N, P_N, A_{\textup{cr}N}\right)$ which achieves an average accuracy of $96(\pm 1)\%$, very close to the highest value. 

\section{\label{sec:L6}Conclusions}

We developed a protocol that integrates image analysis and deep learning  techniques to classify fracture patterns produced with five different solvents, specifically focused on improving the accuracy in determining the solvent used in a desiccation process. First, the fracture patterns were characterized based on morphological features. Then, frequency histograms were generated from a normalized dataset and used to train a large number of MLP neural network variants. We found that 1) the accuracy is improved when the number of neurons decreases progressively in each hidden layer, 2) the crack area is the most important feature in the correct identification, and 3) the accuracy can be enhanced by combining this parameter with other morphological features. Moreover, to optimize the processing time, the four-feature set $\left(\lambda, A_N, P_N,A_{\textup{cr}N}\right)$ is recommended to identify the solvent involved in the desiccation cracking process of starch-liquid slurries, achieving an average accuracy of $96(\pm 1)\%$. These results can motivate further research to determine if current classification methods in different fields of science, engineering and medicine (based on individual parameters) could be improved in their classification accuracy by including the proposed protocol to identify an optimal set of parameters for classification.

\backmatter

%
%
%

\bmhead{Acknowledgments}

The authors gratefully acknowledge financial support from CONACyT Mexico (now SECIHTI) and VIEP-BUAP project 2024 and 2025.  J. I. M. C. acknowledges Ph.D. scholarship 1079089 ("Becas nacionales CONAHCyT 2024").

\section*{Declarations}

\begin{itemize}

\item Funding: Research supported by VIEP-BUAP 2024-2025 (3,000 USD two year total funding) \\

\item Conflict of interest: the authors have no relevant financial or non-financial interests to disclose.\\

\end{itemize}

\bibliographystyle{sn-mathphys} 

\bibliography{references}

\end{document}